\documentclass{ws-procs10x7}

\begin{document}

\title{High $P_T$ multi-lepton production at HERA}

\author{E. Sauvan}

\address{On behalf of the H1 and ZEUS Collaborations\\
Centre de Physique des Particules de Marseille\\
163 Avenue de Luminy F-13288 Marseille cedex 9, France\\E-mail: sauvan@cppm.in2p3.fr}

\twocolumn[\maketitle\abstract{
Multi-electron and multi-muon production at high transverse momentum is measured in $ep$ scattering at HERA. Previous published analyses are extended, combining new HERA II data taken in 2003--2004 with previous HERA I data sample. 
In addition events with high $P_T$ electrons and muons are investigated here for the first time. Yields of di-lepton and tri-lepton events are measured and a general good agreement is found with the Standard Model prediction, dominated by photon-photon interactions. 
Events are observed with leptons of high transverse momenta in a domain where the Standard Model prediction is low.

}]

\section{Multi-lepton processes}

At HERA, two experiments (H1 and ZEUS) study electron-proton collisions with a centre-of-mass energy of up to 320 GeV. The first measurements of multi-electron and multi-muon production at high transverse momentum ($P_T$) in $ep$ collisions have been published recently by the H1 Collaboration\cite{Aktas:2003jg,Aktas:2003sz}.
Preliminary analyses have been caried out by the ZEUS Collaboration\cite{MEZeus}.
A complete preliminary analysis of the multi-lepton topologies  has been performed by the H1 Collaboration including the most recent data~\cite{MlepH1}.
Within the Standard Model (SM) the production of multi-lepton events in $ep$ collisions proceeds mainly via photon-photon interactions in which quasi-real photons radiated from the incident electron and proton interact to produce a pair of leptons: $\gamma \gamma \rightarrow l^+l^-$ \cite{Ver83}. For multi-electron final states, the background comes mainly from neutral current Deep Inelastic Scattering (DIS) or elastic Compton scattering, while for multi-muon final states the background is negligible.

\section{Multi-electron events}

\begin{figure}
\epsfxsize170pt
\figurebox{}{}{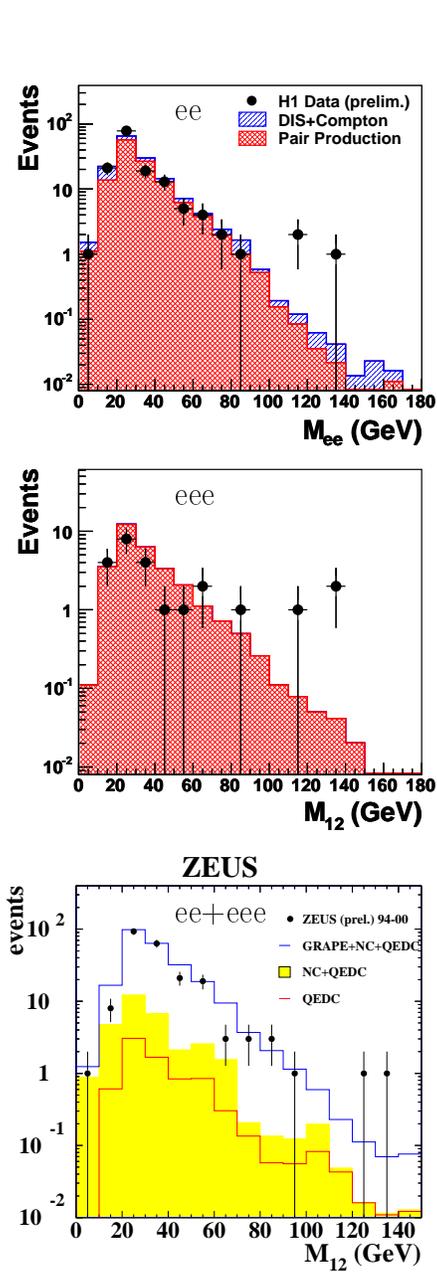}
\caption{Distribution of the invariant mass $M_{12}$ of the two highest $P_T$ electrons for the H1 analysis (two upper figures). Events classified as di-electrons and tri-electrons are shown. The mass distribution from ZEUS analysis with di-electrons and tri-electrons samples combined is displayed in the bottom figure.}
\label{fig:me_h1_zeus}
\end{figure}

The selection of multi-electron events is done by requiring at least two central electrons (20$^\circ$ $< \theta_e$ $<$ 150$^\circ$ for H1 and 17$^\circ$ $< \theta_e$ $<$ 164$^\circ$ for ZEUS) with $P_T^{e_1}(P_T^{e_2}) >$ 10(5) GeV (H1) or $P_T^{e_1}(E^{e_2}) >$ 10(5) GeV (ZEUS). 
An isolated charged track has to be associated with the calorimeter energy deposit. 
A third electron is allowed to be present in a wider angular range (5$^\circ$ $< \theta_e$ $<$~175$^\circ$). 
The selected events are classified by the number of identified electrons into di- and tri-electron samples (ee and eee, respectively). 

The H1 results and the ZEUS measurements are presented in Tab.~\ref{tab:me_h1zeus}. The H1 analysis published in \cite{Aktas:2003jg} is extended here to a higher luminosity combining new HERA II data taken in 2003--2004 ($\cal{L}$ = 45 pb$^{-1}$) with previous HERA I data sample from 1994--2000 ($\cal{L}$ = 118 pb$^{-1}$). 
The di-electron sample is dominated by pair production with only a 15--20\% contribution from other SM processes. 
In the tri-electron sample, the background contribution is negligible. 
Both the H1 and ZEUS observations are in good agreement with the predicted overall yields. The main difference between the H1 and ZEUS predictions for the signal is due to the different angular range allowed for the central electrons.
A few events with invariant mass of the two highest $P_T$ electrons $M_{12} > 100$ GeV are observed in a region in which the SM prediction is low (Fig.~\ref{fig:me_h1_zeus}). Three di-electron events are measured by H1 where 0.44 are expected and 2 by ZEUS for an expectation of 0.77. In the tri-electron sample, 3 events are observed by H1 compared to an expectation of 0.31, while no events are observed by ZEUS where 0.37 are expected. 

The $ep \rightarrow (e)e^+e^- X$ cross section has been measured by H1 \cite{Aktas:2003jg} in a restricted kinematic region where photon-photon processes dominate. The integrated cross section over the considered phase space is measured to be $\sigma$=0.59$\pm$0.08$\pm$0.05 pb, in good agreement with the SM expectation of 0.62$\pm$0.02 pb.

\begin{table*}[t]
\caption{Observed and predicted multi-electron event yields for all selected events and for events with masses $M_{12} > $ 100 GeV. The prediction errors for the H1 analysis include model uncertainties and experimental systematical errors added in quadrature. For the ZEUS analysis, the predicted rates are shown with the statistical errors of the Monte Carlo only.\label{tab:me_h1zeus}}
  \begin{center}
  \begin{tabular}{l c c c c} 
\hline
Selection &  Data & SM & Pair Production & DIS+Compton \\
\hline
\multicolumn{5}{l}{H1 (163 pb$^{-1}$, HERA I+II)}\\
\hline
ee & 147 & 149.8 $\pm$ 24.8 & 125.5 $\pm$ 13.0 & 24.3 $\pm$ 18.7 \\
eee & 24 & 30.4 $\pm$ 3.9 & 30.41 $\pm$ 3.9 & 0.04 $\pm$ 0.06 \\
\hline
ee {\footnotesize ($M_{12}>100$ GeV)} & 3 & 0.44 $\pm$ 0.10 & 0.32 $\pm$ 0.10 & 0.12 $\pm$ 0.03 \\
eee {\footnotesize ($M_{12}>100$ GeV)} & 3 & 0.31 $\pm$ 0.08 & 0.31 $\pm$ 0.08 & ---  \\
\hline
\multicolumn{5}{l}{ZEUS (130 pb$^{-1}$, HERA I)}\\
\hline
ee & 191 & 213.9 $\pm$ 3.9 & 182.2 $\pm$ 1.2 & 31.7 $\pm$ 3.7 \\
eee & 26 & 34.7 $\pm$ 0.5 & 34.7 $\pm$ 0.5 & --- \\
\hline
ee {\footnotesize ($M_{12}>100$ GeV)} & 2 & 0.77 $\pm$ 0.08 & 0.47 $\pm$ 0.05 & 0.30 $\pm$ 0.07 \\
eee {\footnotesize ($M_{12}>100$ GeV)} & 0 & 0.37 $\pm$ 0.04 & 0.37 $\pm$ 0.04 & --- \\
\hline
  \end{tabular}
  \end{center}
\end{table*}

\section{Measurement of multi-muon events} 

The production of muon pairs at high $P_T$ was also studied by the H1 and ZEUS Collaborations.
In the ZEUS preliminary analysis \cite{MuonZeus}, muons are identified using central tracker reconstructed tracks, calorimetric deposits and muon chamber signals. At least two muons with $P_T^{\mu_1,\mu_2} >$ 5~GeV in the angular range 12$^\circ$ $< \theta_\mu$ $<$~157$^\circ$ and with an invariant mass $M_{\mu\mu} >$ 5 GeV are required.

\begin{figure}
\epsfxsize150pt
\figurebox{}{}{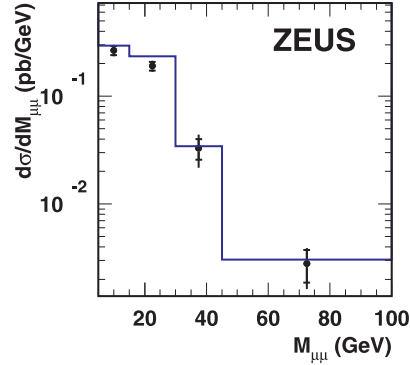}
\caption{Differential muon pair production cross-section as a function of the invariant mass of the muon pair $M_{\mu\mu}$, measured by the ZEUS Collaboration.}
\label{fig:zeusmuon}
\end{figure}

With an analysed data sample corresponding to an integrated luminosity of 101~pb$^{-1}$, 255 multi-muon events are observed.
Figure \ref{fig:zeusmuon} presents the visible cross section measured as a function of the invariant mass of the muon pair compared to the SM prediction; a good agreement is observed.
The integrated cross section in the visible phase space is measured to be 6.17$\pm$0.39(stat.)$^{+0.49}_{-0.43}$(syst.)$\pm$0.12(lumi.) pb, in good agreement with the prediction of 7.13~pb. 

\begin{figure}
\epsfxsize140pt
\figurebox{}{}{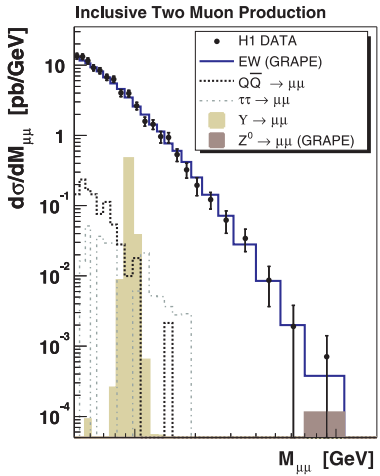}
\caption{Differential muon pair production cross-section as a function of di-muon mass, from the H1 analysis.}
\label{fig:h1muon}
\end{figure}

A similar analysis has been published by H1~\cite{Aktas:2003sz}. The two muons are selected with a minimum $P_T$ of 2 and 1.75 GeV, respectively, in the angular range  20$^\circ$ $< \theta_\mu$ $<$~160$^\circ$. 
The cross section measured as a function of the invariant mass of the muon pair is displayed in Fig. \ref{fig:h1muon} and is in good agreement with the SM prediction. 
The integrated cross section is measured to be 46.4$\pm$1.3$\pm$4.5 pb, also in very good agreement with the prediction of 46.1$\pm$1.4 pb. 
Elastic and inelastic muon pair production processes are distinguished by tagging hadronic activity. In the analysed phase space, integrated cross section of $\sigma^{el}_{\mu\mu}$=25.3$\pm$1.0$\pm$3.5 pb for elastic di-muon production and of $\sigma^{el}_{\mu\mu}$=20.9$\pm$0.9$\pm$3.2 pb for inelastic production are measured. These measurements are in good agreement with the expected cross sections of 24.6$\pm$0.3 pb and 21.5$\pm$1.1 pb, respectively.

\section{Measurement of multi-lepton events} 

Multi-electron and multi-muon measurements have been recently extended by H1 to the $e\mu$ and $e\mu\mu$ topologies and to a higher luminosity, combining new HERA II data taken in 2003--2004 with previous HERA I data sample \cite{MlepH1}. The multi-lepton selection requires that there be at least two central (20$^\circ < \theta < $150$^\circ$) lepton (electron or muon) candidates of which one must have $P_T^l >$ 10 GeV and the other $P_T^l >$ 5 GeV. 
Additional electron candidates are identified in the detector with an energy above 5 GeV in the range 5$^\circ < \theta <$ 175$^\circ$. Additional muons with $P_T > 2$ GeV in the range 20$^\circ < \theta <$ 160$^\circ$ are also looked for. 
The lepton candidates are ordered according to decreasing $P_T$, $P_T^{l_i} > P_T^{l_{i+1}}$.
The selected events are classified in a two lepton sample, if only two central leptons are identified, and in a three lepton sample if exactly one additional lepton candidate is identified. According to the flavour of identified leptons, those samples are further classified into ee, $\mu\mu$, e$\mu$, eee and e$\mu\mu$.

\begin{figure}
\epsfxsize170pt
\figurebox{}{}{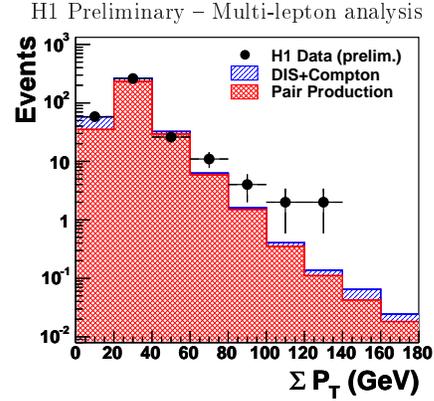}
\caption{Distribution of the scalar sum of $P_T$ of leptons compared to expectations. All di-lepton and three-lepton event classes are combined. The analysed data sample corresponds to an integrated luminosity of 163 pb$^{-1}$ obtained by the H1 experiment.}
\label{fig:SumEt_All_lep}
\end{figure}

The observed event yields in all channels are in good agreement with the SM expectations. In the e$\mu$ and e$\mu\mu$ channels, 86 and 41 events are observed, compared to the SM  predictions of 78.4$\pm$12.0 and 39.5$\pm$6.5, respectively.
The distributions of the scalar sum of $P_T$ of all identified leptons for the combination of di- and three-lepton samples is shown in Fig. \ref{fig:SumEt_All_lep}. For $\sum P_T >$ 100 GeV 4 events are observed while 0.61 $\pm$ 0.11 are expected. These four data events corresponds to the three high mass ee events observed in HERA I data \cite{Aktas:2003jg} and one new e$\mu\mu$ event observed in HERA II data \cite{MlepH1}.

\section{Conclusions} 

All event topologies containing two or three visible leptons (electrons or muons) have been measured in electron-proton collisions. Results on multi-electron and multi-muon events have been published by the H1 Collaboration. 
The ZEUS experiment also presented preliminary results in those channels.
Previous analyses have been recently extended by H1, using new HERA II data taken in 2003--2004 amounting to an integrated luminosity of 45 pb$^{-1}$.
Good overall  agreement with the Standard Model prediction is observed. 
The differential cross sections of the production of electron and muon pairs have been measured and found in very good agreement with the Standard Model. 
However, four events are also observed with a scalar sum of lepton transverse momenta greater than 100 GeV, compared to a Standard Model prediction of 0.61 $\pm$ 0.11.

\end{document}